\documentclass[twocolumn,twoside]{IEEEtran}

\ifCLASSINFOpdf

\else

\fi

\ifCLASSOPTIONcompsoc
    \usepackage[caption=false, font=normalsize, labelfont=sf, textfont=sf]{subfig}
\else
    \usepackage[caption=false, font=normalsize]{subfig}
\fi
\usepackage{lipsum}%
\usepackage[dvipsnames]{xcolor}
\usepackage{algorithm,algorithmic}
\usepackage{balance}
\usepackage{multicol}   % for equalize of last page columns1	
\usepackage{cite}
\usepackage{gensymb}
\usepackage{multirow}
\usepackage{graphics}
\usepackage{epsfig}
\usepackage{graphicx}
\usepackage{epstopdf}
\usepackage{textcomp}
\usepackage{amsmath}
\usepackage{mathtools}
\usepackage{filecontents}
\usepackage{lipsum,color}
\usepackage{amssymb}
\usepackage{float}
\usepackage{colortbl} % for painting tables
\usepackage{times} % assumes new font selection scheme installed
\usepackage{amsthm}  % for proof
\usepackage{amsfonts}

\theoremstyle{break}
\begin{document}
\title{Optimizing UAV Trajectory for Emergency Response Operations under Real 3D Environments: Integrating Priority Levels and LoS Constraints}
%\title{UAV Trajectory Optimization for Emergency Response: Integrating THz Communication and 3D Urban Navigation}

\author{ Mohammad~T.~Dabiri,~Mazen~Hasna,~{\it Senior Member,~IEEE},~Saud~Althunibat,~{\it Senior Member,~IEEE}, \\~and
	~Khalid Qaraqe,~{\it Senior Member,~IEEE}
\thanks{This publication was made possible by NPRP14C-0909-210008 from the Qatar National Research Fund (a member of The Qatar Foundation). The statements made herein are solely the responsibility of the author[s].}
\thanks{Mohammad Taghi Dabiri, and Mazen Hasna are with the Department of Electrical Engineering, Qatar University, Doha, Qatar.  (E-mail: m.dabiri@qu.edu.qa; hasna@qu.edu.qa).}
\thanks{Saud Althunibat is with the Department of Communications Engineering, Al-Hussein Bin Talal University, Jordan (E-mail: saud.althunibat@ahu.edu.jo).}
\thanks{Khalid A. Qaraqe is with the Department of Electrical and Computer Engineering, Texas A$\&$M University at Qatar, Doha 23874, Qatar (E-mail: khalid.qaraqe@qatar.tamu.edu).}
}

\maketitle
%%%%%%%%%%%%%%%%%%%%%%%%%%%%%%%%%%%%%%%%%%%%%%%%%%%%%%%%%%
%%%%%%%%%%%%%%%%%%%%%%%%%%%%%%%%%%%%%%%%%%%%%%%%%%%%%%%%%%
\begin{abstract}
%%%%%%%%%%%%%%%%%%%%%%%%%%%%%%%%%%%%%%%%%%%%%%%%%%%%%%%%%%
%%%%%%%%%%%%%%%%%%%%%%%%%%%%%%%%%%%%%%%%%%%%%%%%% 
Unmanned Aerial Vehicles (UAVs) have emerged as a critical component in next-generation wireless networks, particularly for disaster recovery scenarios, due to their flexibility, mobility, and rapid deployment capabilities. This paper focuses on optimizing UAV trajectories to ensure effective communication in disaster-stricken areas using terahertz (THz) links. We address specific challenges such as energy consumption, user priority levels, and navigating complex urban environments to maintain Line of Sight (LoS) connections amidst 3D obstacles. Our contributions include the development of a detailed modeling approach using online 3D map data, the formulation of an optimal trajectory optimization problem, and the proposal of a Genetic Algorithm (GA)-based method alongside an enhanced heuristic algorithm for faster convergence. Through 3D simulations, we demonstrate the trade-off between minimizing total service time and prioritizing higher-weight nodes, showing the impact of different priority weight factors on the trajectory time. The proposed algorithms are evaluated using real-world data from the West Bay area of Doha, Qatar, demonstrating their effectiveness in optimizing UAV trajectories for emergency response. 
\end{abstract}
\IEEEpeerreviewmaketitle

%%%%%%%%%%%%%%%%%%%%%%%%%%%%%%%%%%%%%%%%%%%%%%%%%%%%%%%%%%%%
%%%%%%%%%%%%%%%%%%%%%%%%%%%%%%%%%%%%%%%%%%%%%%%%%%%%%%%%%%%%
\section{Introduction}
%%%%%%%%%%%%%%%%%%%%%%%%%%%%%%%%%%%%%%%%%%%%%%%%%%%%%%%%%%%%
%%%%%%%%%%%%%%%%%%%%%%%%%%%%%%%%%%%%%%%%%%%%%%%%%%%%%%%%%%%% 

\IEEEPARstart{U}{nmanned} Aerial Vehicles (UAVs) have emerged as a key component in next-generation wireless networks, owing to their flexibility, mobility, and rapid deployment capabilities. These attributes make UAVs particularly valuable for enhancing communication networks by providing Line of Sight (LoS) connections, which is crucial for high-frequency millimeter-wave (mmWave), terahertz (THz), and free-space optical (FSO) links \cite{dabiri2022modulating}. The integration of UAVs in network infrastructure promises improvements in data rates, spectrum efficiency, low latency, and overall network reliability. In disaster recovery scenarios, the requirements for communication networks significantly differ from normal conditions. Rapid and reliable re-establishment of communication links is vital for coordinating emergency response and aiding affected populations. UAVs are uniquely suited for these situations, as they can be quickly deployed to create or restore communication networks over disrupted areas.

To this end, the primary focus of this paper is on optimizing UAV trajectories to ensure effective communication in disaster recovery scenarios. This involves addressing specific challenges such as energy consumption, prioritizing users based on urgency, and navigating complex urban environments to maintain LoS connections in the presence of real 3D obstacles and buildings. By concentrating on these aspects, this study aims to enhance the efficiency and reliability of UAV-assisted communication networks during critical disaster recovery operations.

%zeng20243d
%--------------------------
%--------------------------
\subsection{Literature Review}
%--------------------------
%--------------------------
Due to their high altitude and maneuverability, UAVs are more likely to establish LoS links, thereby enabling higher capacity networks compared to terrestrial links. Numerous studies have been conducted on UAV-based networks, exploring various aspects such as LoS probability, energy efficiency, and deployment strategies \cite{8247211,8918497,9457160,9800925,zeng20243d}. However, most of these works focus on LoS probability analysis, which may not be adequate for high-frequency directional THz/FSO links, as these signals cannot penetrate obstacles, making exact LoS determination crucial.

Gemmi \textit{et al.} \cite{gemmi2022properties,gemmi2022cost} proposed GPU-based algorithms using 3D digital maps to find optimal antenna placements on building corners to maximize LoS coverage. These methods, though effective for terrestrial networks, are not directly applicable to UAV-based systems. Other studies have examined UAV networks in the presence of 3D obstacles, employing techniques such as ray tracing \cite{hu2020low}, binary channel concepts \cite{9709500}, and geometric analysis \cite{li2022geometric,yi2022joint,yi20233,tang2021performance,zhu2022geometry} to model and mitigate LoS blockages.

In \cite{hu2020low}, a UAV-aided relay network is considered using a ray-tracing model to provide coverage for ground users in urban areas with numerous obstructions. The concept of binary channels for energy delivery is introduced in \cite{9709500}, which specifies the LoS state of users. The idea of channel knowledge maps for environment-aware wireless communications is discussed in \cite{zeng2021toward}, and a coverage-aware navigation approach is proposed in \cite{zeng2021simultaneous} to achieve ubiquitous 3D communication coverage for UAVs.

To address the challenges posed by 3D obstacles, several works have focused on UAV positioning and placement. For instance, \cite{wang2020placement} and \cite{sabzehali20213d} explore 3D placement of UAVs to provide LoS coverage for users, using simplified models that may not capture the full complexity of real environments. In contrast, more detailed geometric-based LoS models are proposed in \cite{9044827}, which have inspired subsequent research \cite{lin2021adaptive,xiao2023joint,dabiri2024coverage} to accurately model and navigate 3D urban landscapes.

In the context of UAV deployment for emergency networking, Xiao \textit{et al.} \cite{xiao2023joint} present a joint optimization framework considering positioning of UAVs. Similarly, in \cite{dabiri2024coverage}, the authors propose adaptive UAV deployment strategies focusing on maximizing coverage and ensuring reliable communication. Both studies primarily focus on positioning UAVs in the presence of 3D obstacles to ensure LoS connectivity for a distributed number of users among these real obstacles.

In post-disaster scenarios, network congestion is a common issue due to increased demand for communication. To address this, higher frequency bands, such as THz frequencies, are ideal as they are less likely to experience congestion. THz frequencies have several advantageous properties: they allow for the use of small, high-gain antenna arrays on UAVs, which is crucial given the weight constraints of UAVs. The high gain of these antenna-arrays improves the signal-to-noise ratio and reduces beamwidth, thus minimizing interference and preventing congestion. However, a critical challenge with THz signals is their inability to penetrate 3D obstacles. Therefore, ensuring that users are in the LoS of the UAV during service is essential. Unlike probabilistic LoS analysis, precise trajectory planning of UAVs is required, taking into account the physical characteristics of 3D obstacles and tall buildings.
Additionally, in emergency situations, it is crucial to provide prioritized service to users based on urgency. Users in critical conditions should be served first with a higher QoS. This necessity for prioritization and ensuring high QoS for critical users, along with guaranteeing LoS during service delivery to distributed users among 3D obstacles and prioritizing users during trajectory planning, are aspects that have not been thoroughly explored in the existing literature.

%% -------------------------
%% -------------------------
\subsection{Contribution}
%% -------------------------
%% -------------------------
In this paper, we consider a disaster recovery scenario where UAVs are used to manage and provide services to users trapped in crisis areas using THz links until rescue forces arrive. To this end, we aim to optimize the UAV trajectory in a real 3D environment with obstacles, ensuring that users remain in the LoS of the UAV during service. Additionally, the trajectory planning is done based on user priority levels, ensuring that higher priority users receive service first. The UAV's trajectory is planned to minimize the total mission time, including moving from the charging station, providing emergency alerts to all users, and returning to the charging station while considering the UAV's energy consumption constraints. To the best of the authors' knowledge, this issue has not been the subject of any existing work in the literature.

The primary contributions of this paper are as follows:
\begin{itemize}
	\item We present a detailed modeling of 3D obstacles using online 3D map data and propose an algorithm to rapidly determine the LoS status for any UAV position in a 3D space. This approach ensures that the UAV can navigate through complex urban environments, maintaining reliable communication links with users.
	\item An optimal trajectory optimization problem is formulated to minimize the total mission time while integrating priority levels of users and LoS constraints.
	\item We develop a Genetic Algorithm (GA)-based method to solve the optimization problem, considering the complexities of continuous and discrete decision variables.
	\item An enhanced heuristic algorithm is proposed to achieve faster convergence by leveraging environmental physical information and user priority weights.
	\item Through 3D simulations, we demonstrate the trade-off between minimizing total service time and prioritizing higher-weight nodes, showing the impact of different priority weight factors on the trajectory time.
	\item The proposed algorithms are evaluated using real-world data from the West Bay area of Doha, Qatar, demonstrating their effectiveness in optimizing UAV trajectories for emergency response.
\end{itemize}

%%%%%%%%%%%%%%%%%%%%%%%%%%%%%%%%%%%%%%%%%%%%%%%%%%%%%%%%%%%%%%%%
%%%%%%%%%%%%%%%%%%%%%%%%%%%%%%%%%%%%%%%%%%%%%%%%%%%%%%%%%%%%%%%% VERSUS P_T
\begin{figure}
	\begin{center}
		\includegraphics[width=3.4 in]{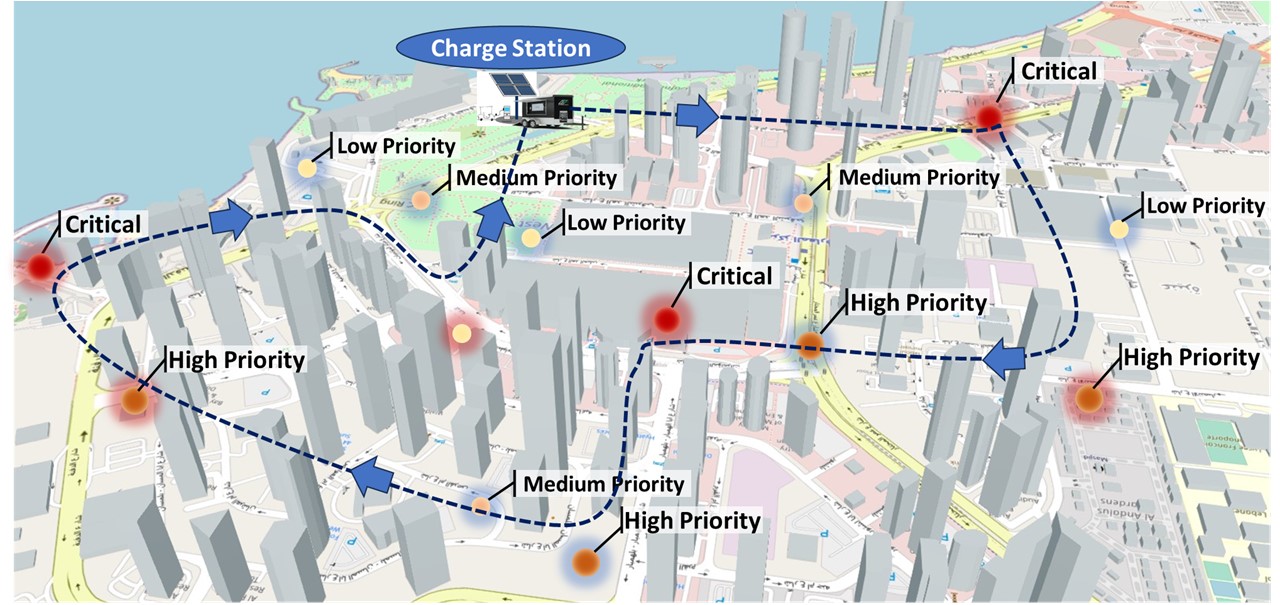}
		\caption{A graphical illustration of distributed users among 3D obstacles in the West Bay area of Doha, Qatar, where each user is assigned a priority level. The UAV starts from the charging station, provides services to users based on their priority levels, and returns to the charging station in the shortest possible time.}
		%\caption{Graphical example of Doha, Qatar, West Bay towers where the dominant 3D obstacles are cube-shaped buildings \cite{dOHA_WE}.}
		\label{sm1}
	\end{center}
\end{figure}
%%%%%%%%%%%%%%%%%%%%%%%%%%%%%%%%%%%%%%%%%%%%%%%%%%%%%%%%%%%%%%%%
%%%%%%%%%%%%%%%%%%%%%%%%%%%%%%%%%%%%%%%%%%%%%%%%%%%%%%%%%%%%%%%%
%

\begin{table*} %field of view comparision can be good
	\caption{ UAV Trajectory and Communication Planning with Priority-Based User Locations} % title of Table
	\centering % used for centering table
	\begin{tabular}{c } % centered columns (3 columns)
		\includegraphics[width=7 in]{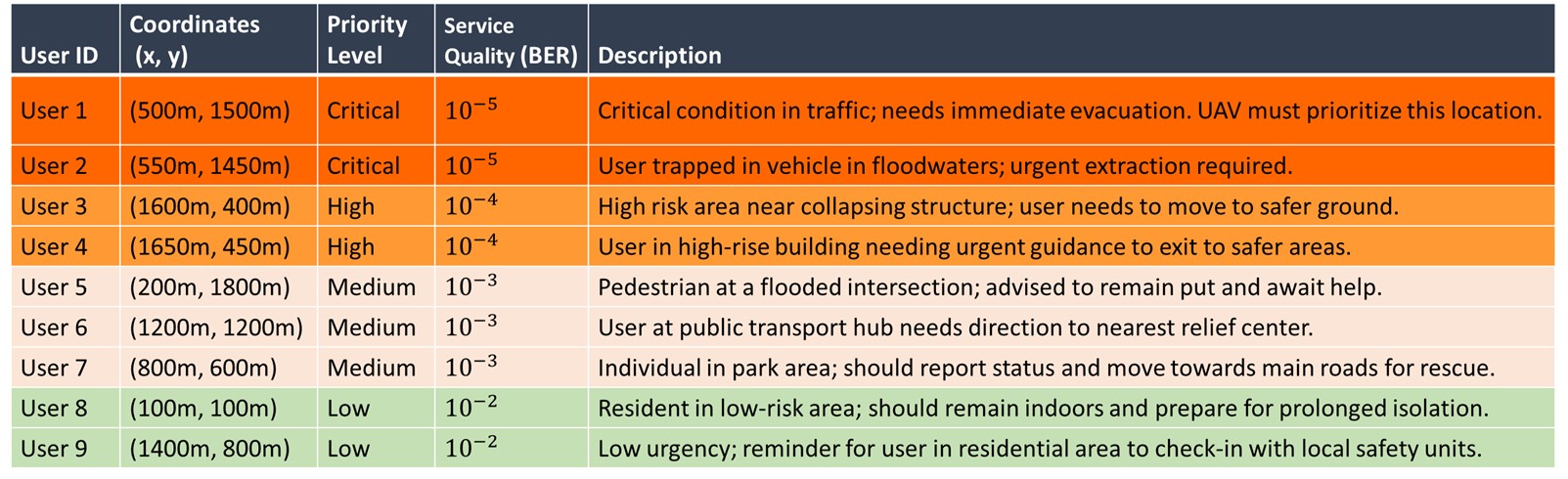}
	\end{tabular}
	\label{Tab_prit} 
\end{table*}

\section{System Model and Optimization Problem}

\subsection{Problem Definition}
During a disaster, RF may experience congestion, but THz frequencies, due to their smaller wavelengths, which enable the use of compact antenna arrays to reduce the effects of interference and network congestion. However, these communications face the challenge of requiring a LoS path, as THz signals cannot penetrate 3D obstacles \cite{dabiri2022general,dabiri2023optimal}.

The problem is to optimize the trajectory of a UAV deployed in a disaster-stricken area to facilitate effective and timely communication under real 3D obstacles. To this end, we consider a scenario where ground users are randomly distributed in disaster-stricken areas, denoted as locations $\mathbf{r}_{i} (x_i, y_i, z_i) $, where $i\in\{1,...,N_p\}$ and $N_p$ is the total number of users. A UAV is tasked with departing from a charging station and, after providing service to all users in the minimum possible time, returning back to the charging station. The objective is to minimize the overall mission time while ensuring that communications are reliably delivered to various users based on their priorities and specific needs. Additionally, the UAV must navigate through 3D obstacles/buildings to ensure LoS connectivity, which complicates the trajectory planning. To guarantee LoS during the trajectory, accurate 3D city data is required, which can be obtained from the following online sources:
\begin{itemize}
\item 3D City Database (3DCityDB): This database provides a robust platform for storing, managing, and visualizing virtual 3D city models based on the CityGML standard \cite{3dcitydb}. It supports complex analyses and integrates with various cloud services, exporting data in formats like KML, COLLADA, and glTF.

\item Open City Model: This resource compiles open CityGML data for buildings in the United States by integrating multiple datasets \cite{opencitymodel}. It offers detailed 3D building geometries and attributes such as building footprints, height, and coordinates. 

\item CADMAPPER: This tool converts data from sources like OpenStreetMap, NASA, and USGS into organized CAD files. It is particularly useful for designers, providing detailed building height and topography data \cite{cadmapper}. 

\item Blender-OSM: This site provides map data extracted from OpenStreetMap (OSM) and made available through Blender \cite{blender-osm}. It offers comprehensive 3D city data suitable for urban modeling and visualization. 
\end{itemize}

\subsection{A Practical Realization}
As a practical example, Table \ref{Tab_prit} details the UAV deployment strategy in an urban disaster scenario after a major flood, where each user is assigned a location, a priority level, and a service quality defined by Bit Error Rate (BER). UAVs play a pivotal role in establishing immediate and reliable communication with individuals in disaster areas, prioritizing them both in terms of communication quality and service timing until rescue teams can intervene. The strategic deployment plan outlined prioritizes users based on the urgency of their circumstances, ensuring that those in critical conditions, who may be trapped or in imminent danger, are serviced first and provided with the most reliable communication channels, characterized by the lowest BER ($10^{-5}$). This approach ensures that vital lifesaving information reaches them without delay.

%% ---------------------------------
%% ---------------------------------
\subsection{Optimization Problem Formulation}
%% ---------------------------------
%% ---------------------------------

\subsubsection{Objective Function}
The objective is to minimize the total mission time:
\begin{align}
	\label{eq:objective_min_time}
	\text{Minimize} \quad T = t_\text{end} 
\end{align}
where \( t_\text{end} \) is the time at which the UAV starts serving the last user.%, 

%------------------
\subsubsection{Constraints}
%-------------------
Priority servicing constraint can be formulated as
\begin{align}
	\label{eq:priority_constraint}
	w_i > w_j \Rightarrow t_i < t_j
\end{align}
where $t_i$ is the time at which the UAV starts serving user \( i \), and $0<w_i\leq1$ is the priority level of user $i$. 
This ensures that users with higher priority (higher \( w_i \)) are served before those with lower priority (lower \( w_i \)).

%Energy Consumption Constraint}:
To ensure that the UAV's energy consumption does not exceed the available energy capacity, we have
\begin{align}
	\label{eq:energy_constraint}
	E_{\text{total}} \leq E_{\text{max}}
\end{align}
where \( E_{\text{total}} \) is the total energy consumed by the UAV, which can be approximated by:
\begin{align}
	E_{\text{total}} \approx P_{\text{UAV}} \cdot \left( t_{N_p} + t_{\text{return}} \right)
\end{align}
where \( P_{\text{UAV}} \) is the power consumption of the UAV, and \( t_{\text{return}} \) is the time required for the UAV to return back to the charging station from the location of last user. This condition ensures that the UAV has enough energy to complete its mission and return to the charging station.

%\textbf{Quality of Service (QoS) Constraint}:
During its trajectory, the UAV must guarantee the quality of service (QoS) required by each user.
The quality of service for each user, measured by the BER, must be proportional to the user's priority level \( w_i \). This ensures that users with higher priority (higher \( w_i \)) are provided with better QoS (lower BER):
\begin{align}
	\ln(\text{BER}_i ) \leq \ln( \text{BER}_\text{th,1} ) \cdot (1 - w_i) + \ln( \text{BER}_\text{th,0} ) \cdot w_i
	\label{eq:ber_wi}
\end{align}
where \( \text{BER}_\text{th,1} \) is the maximum acceptable BER or threshold BER for $w_i=1$ and \( \text{BER}_\text{th,0} \) is the threshold BER for $w_i=0$. Note that $\text{BER}_\text{th,1}<<\text{BER}_\text{th,0}$.

Given the inability of THz signals to penetrate 3D obstacles, the primary condition for providing service to the user is that the UAV must place the user in a LoS state.
To ensure that the UAV maintains a direct LoS with each user during their communication time, the UAV must be positioned such that there are no obstacles obstructing the direct path between the UAV and the user.
For the UAV to be in LoS with user \( i \) at time \( t_i \), the following condition must be satisfied:
\begin{align}
	\text{LoS}_{i,t_i} = 
	\begin{cases}
		1, & \text{if no obstacles intersect the line segment} \\ & \text{between}~\mathbf{r}_{u,t_i}~\text{and}~\mathbf{r}_i \\
		0, & \text{otherwise}
	\end{cases}
	\label{eq:los_condition}
\end{align}
where $\mathbf{r}_{u,t} = (x_{\text{u}}(t), y_{\text{u}}(t), z_{\text{u}}(t))$ is the UAV's position at time $t$.

Before solving this optimization problem, we will examine constraints \eqref{eq:ber_wi} and \eqref{eq:los_condition} in more detail in the next section.

\subsection{Optimization Problem Complexity}
The optimization problem presented in this study is highly complex due to the practical constraints incorporated. For instance, ensuring LoS and maintaining a high QoS require precise knowledge of the coordinates and physical characteristics of all obstacles within the environment. This introduces a significant computational complexity as the UAV's trajectory must be continually adjusted to navigate around these obstacles while maintaining LoS with the users and meeting the QoS requirements. Additionally, discrete decision variables are involved in determining the LoS status, the service order of the users, and resource allocation and scheduling.

Therefore, this optimization problem is NP-hard due to the combination of continuous and discrete decision variables, making it a mixed-integer optimization problem. The necessity to maintain LoS and QoS, coupled with priority-based service and energy constraints, further complicates the solution space. Given the complexities of the optimization problem, traditional optimization methods may fall short in efficiently solving this problem. As a result, the problem requires sophisticated approaches that can handle its inherent complexity and provide near-optimal solutions within a reasonable time frame.

	\begin{table}[h]
	\centering
	\caption{Variable Definitions}
	\begin{tabular}{|c|p{6.2cm}|}
		\hline
		\textbf{Notation} & \text{Description} \\ \hline
		$\mathbf{r}_{u,t}$ & $ = (x_{\text{u}}(t), y_{\text{u}}(t), z_{\text{u}}(t))$, the UAV's position at any given time \( t \) \\ \hline
		$\mathbf{r}_\text{station}$ &=\(\mathbf{r}_{u,0} = (x_{\text{u}}(0), y_{\text{u}}(0), z_{\text{u}}(0))\): Coordinates of charging station \\ \hline
		$\mathbf{r}_{i}$ & = \( (x_i, y_i, z_i) \), the position coordinates of user \( i \) \\ \hline
		$\mathbf{r}'_{nj}$ & \( (x'_{nj}, y'_{nj}, z'_{nj}) \), coordinates of $j$th upper vertices of \(n\)-th cuboid for $j\in\{1,...,4\}$ \\ \hline
		\( t_i \) & The time at which the UAV starts serving user \( i \) \\ \hline
		\( w_i \) & Risk level or priority level of the \(i\)-th user, with a higher value indicating higher priority \\ \hline
		\(I_w\) & Used in objective function \eqref{eq:new_objective} that determines the trade-off between time and priority, where \(I_w \geq 0\) \\ \hline
		\( \text{LoS}_{i,t_i} \) & A binary variable indicating the LoS status between $\mathbf{r}_{u,t_i}$ and $\mathbf{r}_{i}$ at time \( t_i \), where 1 represents a clear LoS and 0 represents an obstructed view \\ \hline
		\( N_p \) & The total number of users \\ \hline
		\( E_{\text{max}} \) & The maximum energy available for the UAV \\ \hline
		\( v_{\text{UAV}} \) & The constant speed of the UAV \\ \hline
		\( d_{\text{total}} \) & The total distance traveled by the UAV \\ \hline
		\( t_{\text{return}} \) & The time required for the UAV to return to the charging station from the last user \\ \hline
		\( E_{\text{total}} \) & The total energy consumed by the UAV \\ \hline
		\( P_{\text{UAV}} \) & The power consumption of the UAV \\ \hline
		\( \text{BER}_i \) & The BER threshold of the \(i\)-th user, which is a function of \( w_i \) \\ \hline
		\( \text{BER}_\text{th,1} \) & Maximum BER threshold for \( w_i = 0 \) \\ \hline
		\( \text{BER}_\text{th,0} \) & Minimum BER threshold for \( w_i = 1 \) \\ \hline
		\( d_{t_i} \) & Distance between $\mathbf{r}_{u,t_i}$ and $\mathbf{r}_{i}$ \\ \hline
		\( d_{\max} \) & Maximum distance at which the BER is \( \text{BER}_\text{th,1} \) when \( w_i = 0 \) \\ \hline
		\( N_b \) & Total number of cuboids representing the obstacles \\ \hline
		%%----------- FOR Section III
		\(\mathbf{S}_{u,i}\) & Matrix representing the grid-based service area for user \(i\) \\ \hline
		\(s_{u,i}(m_x, m_y)\) & Element of the matrix \(\mathbf{S}_{u,i}\) indicating if the grid point is valid (1) or not (0) \\ \hline
		\(M_x\) & Number of grid points in the x direction \\ \hline
		\(M_y\) & Number of grid points in the y direction \\ \hline
		%% ---------- FOR OPtimization Section --
		\( T_{\text{max}} \) & Maximum allowable flight time \\ \hline
		\( P_{\text{UAV}} \) & Power consumption of the UAV \\ \hline
		\( t_{\text{return}} \) & Time required for the UAV to return to the charging station from the last user, defined by \(\frac{\|\mathbf{r}_{u,t_{N_p}} - \mathbf{r}_{\text{station}}\|}{v_{\text{UAV}}}\) \\ \hline
		\( \mathbf{r}_{\text{station}} \) & Coordinates of the charging station \\ \hline
		\( \lambda \) & Penalty factor for exceeding the maximum allowable flight time \\ \hline
		\( \mu \) & Penalty factor for not satisfying the service area constraint \\ \hline
	\end{tabular}
	\label{tab:variables}
\end{table}

%% ----------------------------
%% ----------------------------	
\section{Service Points Characteristics and Requirements}
%% ----------------------------
%% ----------------------------
To ensure effective service delivery by the UAV to the users located at positions $\mathbf{r}_{i} (x_i, y_i, z_i) $ within a 3D obstacle-laden environment, constraints \eqref{eq:ber_wi} and \eqref{eq:los_condition} must be satisfied.
In this section, the modeling and more detailed analysis of constraints \eqref{eq:ber_wi} and \eqref{eq:los_condition} is provided.

\subsection{LoS Condition}
For the UAV to serve the \(i\)-th user from a service point $\mathbf{r}_{u,t_i} = (x_{\text{u}}(t_i), y_{\text{u}}(t_i), z_{\text{u}}(t_i))$, the user must be within the LoS of the UAV as defined in \eqref{eq:los_condition}.

\subsubsection{Modeling Obstacles Using Cuboids}
\begin{figure}
	\centering
	\subfloat[] {\includegraphics[width=3.2 in]{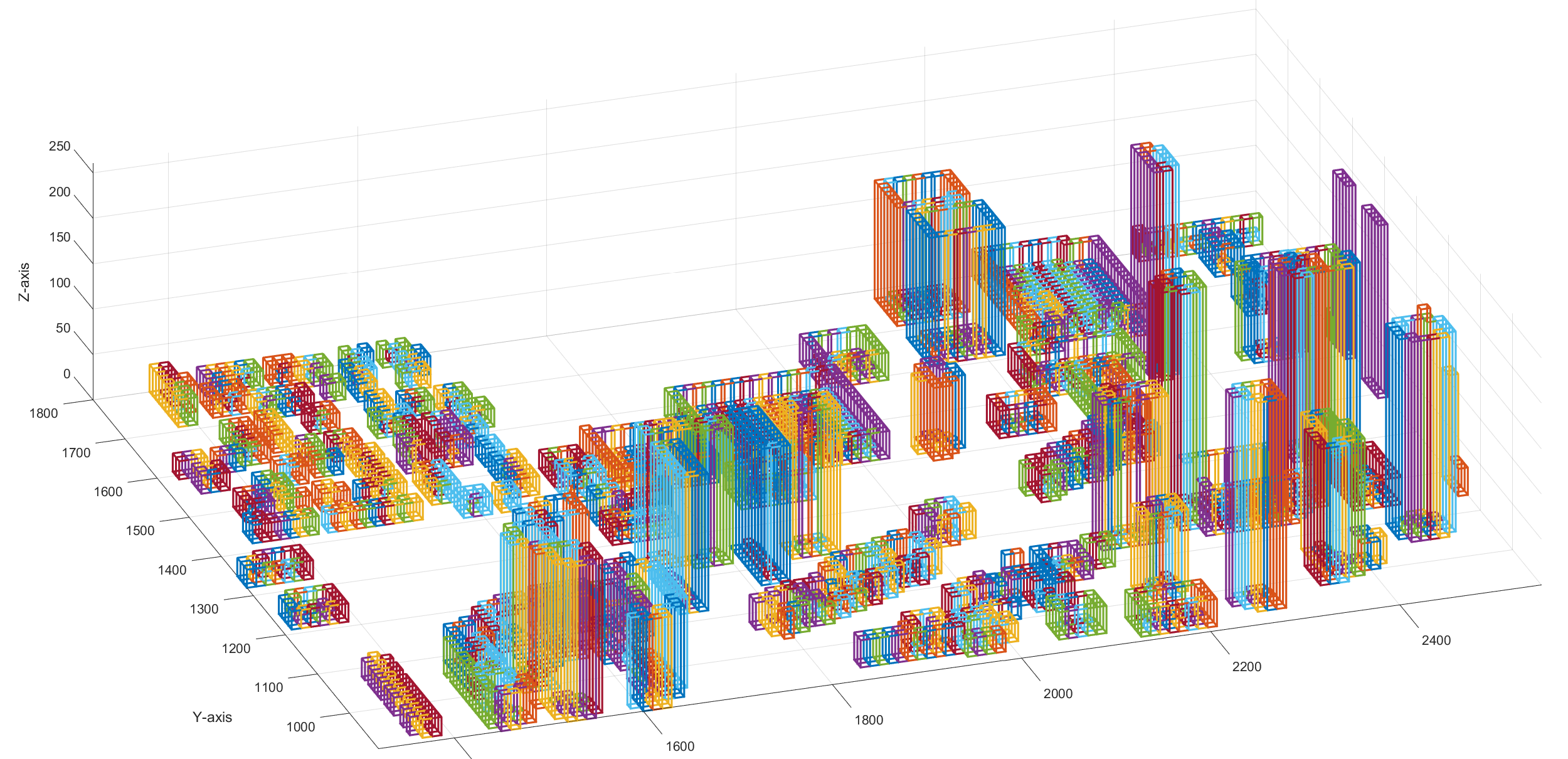}
		\label{cm1}
	}
	\hfill
	\subfloat[] {\includegraphics[width=3.2 in]{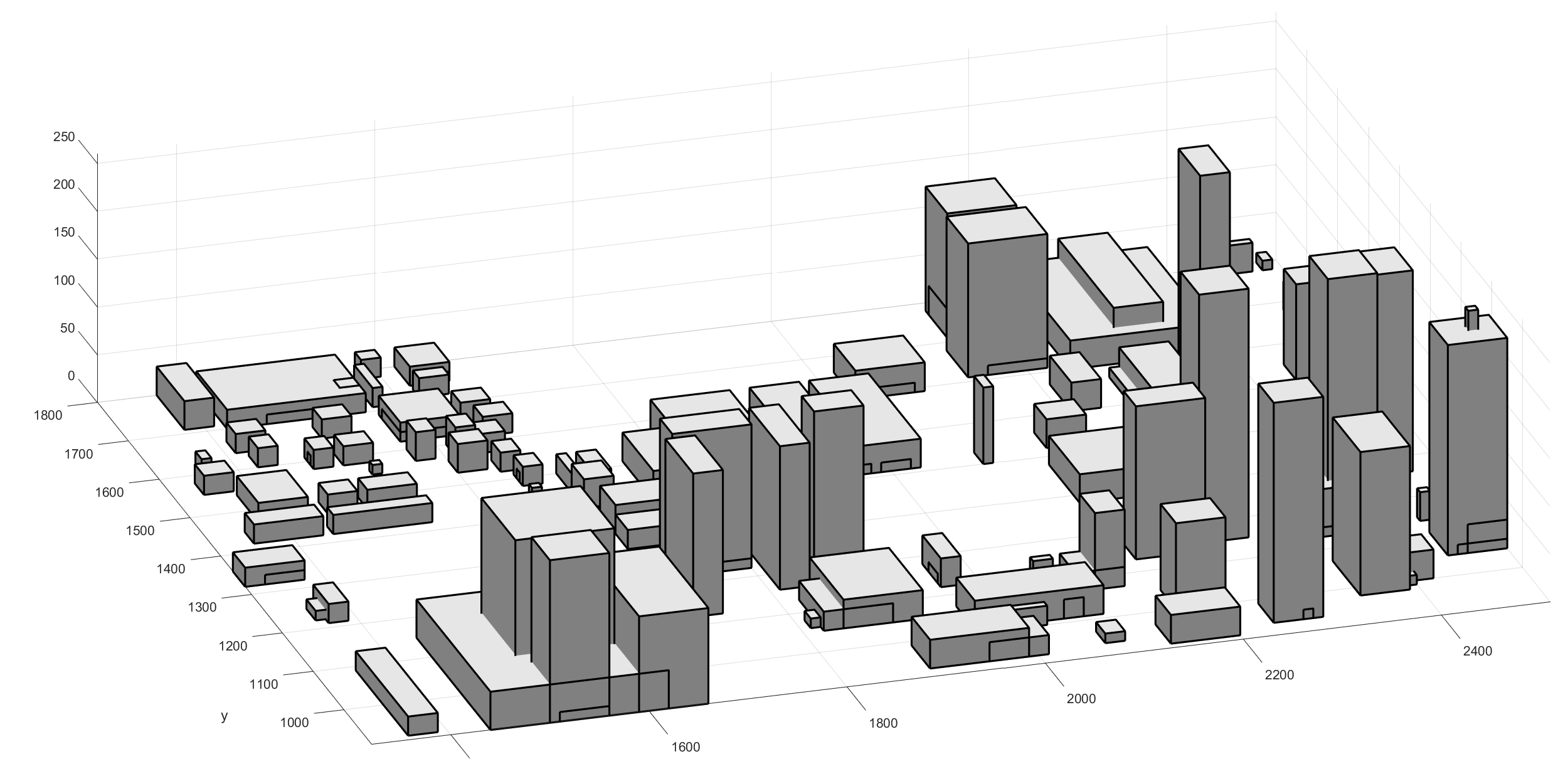}
		\label{cm2}
	}
	\caption{Illustration of the West Bay area in Doha, Qatar, using Blender-OSM data. (a) Shows the urban environment modeled with 3150 cubes to represent tall buildings. (b) The same area after applying an initial filtering to reduce the number of cubes to 128, significantly decreasing the computational load for optimization processes.}
	\label{cm}
\end{figure}

In urban environments, the dominant obstacles are often tall buildings. These buildings can be modeled using cuboids, and the necessary data about them are typically available from online mapping services \cite{3dcitydb,opencitymodel,cadmapper,blender-osm}. 
For example, in Fig. \ref{cm1}, a 3D output of Blender-OSM in the city of Doha, Qatar, West Bay area, is shown, which is estimated with 3150 cubes to represent the urban 3D environment.
Each cuboid is characterized by its four upper vertices in the 3D space. For a given cuboid \( n \) (where \( n \in \{1, \ldots, N_b\} \), and $N_b$ is the number of cubes), the four upper vertices are:
\begin{align}
[ \mathbf{r}'_{n1}, \mathbf{r}'_{n2}, \mathbf{r}'_{n3}, \mathbf{r}'_{n4} ]
\label{eq:cuboid_vertices} 
\end{align}
where $\mathbf{r}'_{nj} = (x'_{nj}, y'_{nj}, z'_{nj}) $.
The faces of the cuboid are typically aligned with the \( xz \)-plane and \( yz \)-plane.

%%%%%%%%%%%%%%%%%%%%%%%%%%%%%%%%%%%%%%%%%%%%%%%%%%%%%%%%%%%%%%%% VERSUS W_Z
%%%%%%%%%%%%%%%%%%%%%%%%%%%%%%%%%%%%%%%%%%%%%%%%%%%%%%%%%%%%%%%%
%%%%%%%%%%%%%%%%%%%%%%%%%%%%%%%%%%%%%%%%%%%%%%%%%%%%%%%%%%%%%%%%  
%

In urban environments, the predominant shape of obstacles is tall buildings, which generally have a cubic appearance or can be modeled with fewer cubes. However, in online map data, depending on the grid resolution, a building might be modeled with several hundred cubes, while it could be represented with just one or a few cubes. Therefore, to reduce computational load, an initial filtering of the cubes in the online maps is necessary. For instance, adjacent cubes of the same height can be approximated by a larger single cube. To achieve a better view, by applying an initial filter to the data in Fig. \ref{cm1}, consisting of $N_b=3150$ cubes, we achieved an approximate representation in Fig. \ref{cm2} that models the same environment with $N_b=128$ blocks. This reduction in $N_b$ significantly reduces the computational load in the optimization process.

%---------------------------------
\subsection{Checking LoS with Cuboids}
%------------------------------------
To determine if the UAV at $\mathbf{r}_{u,t_i} = (x_{\text{u}}(t_i), y_{\text{u}}(t_i), z_{\text{u}}(t_i))$ has a clear LoS to the user at $\mathbf{r}_{i} (x_i, y_i, z_i) $, the line connecting these two points must not intersect any of the cuboid faces in the environment. 
Let us to define the spatial vector \(\mathbf{v}\) connecting the UAV at \(\mathbf{r}_{u,t_i}\) and the user at \(\mathbf{r}_{i}\) as:
\begin{align} \label{spat_vec}
	\mathbf{v_{t_i}} = (x_i - x_{\text{u}}(t_i), y_i - y_{\text{u}}(t_i), z_i - z_{\text{u}}(t_i)).
\end{align}
Each building consists of four side faces that must be checked to ensure that the spatial vector $\mathbf{v_{t_i}}$ does not intersect any of these faces. To verify this, we perform the following steps.

Each cuboid is defined by its four upper vertices as shown in \eqref{eq:cuboid_vertices}.
The four side faces of the cuboid are aligned with the coordinate planes:
\begin{align}
	\begin{cases}
		\text{Two faces in the \( x-z \) plane}, \\ 
		\text{Two faces in the \( y-z \) plane}.
	\end{cases}
\end{align}
Let's denote the two faces on the $x-z$ plane as numbers 1 and 3, and the two faces on the $y-z$ plane as numbers 2 and 4.
To determine if the line segment $\mathbf{v_{t_i}}$ intersects a face in the \( x-z \) plane (where \( y=y'_{nj} \) is constant for $j\in\{1,3\}$), $y'_{nj}$ can be written as a linear combination of $y_{\text{u}}(t_i)$ and $y_i$ as: 
\begin{equation}
	y'_{nj} = (1 - b) y_{\text{u}}(t_i) + b 
	\label{eq:constant_y}
\end{equation}
By rearranging \eqref{eq:constant_y}, we have:
\begin{equation}
	b = \frac{y'_{nj} - y_{\text{u}}(t_i)}{y_i - y_{\text{u}}(t_i)}
	\label{eq:b_y}
\end{equation}
If $b$ is greater than 1 or less than 0, then we can definitively say that there is no intersection between the line segment and the specified plane in the $x-z$ plane.
However, if \( b \) is within the interval \([0, 1]\), there is {\bf a possibility} that the line segment intersects the face. Now, to ensure an intersection, it must be confirmed that \( x \) and \( z \) values are within the face boundaries. To this end, we obtain $x$ and $z$ as:
\begin{equation}
	x(b) = (1 - b) x_{\text{u}}(t_i) + b x_i
	\label{eq:x_b}
\end{equation}
\begin{equation}
	z(b) = (1 - b) z_{\text{u}}(t_i) + b z_i
	\label{eq:z_b}
\end{equation}
Now, for the intersection to occur, the following two conditions must also be met:
\begin{equation}
	\min(x'_{n1}, x'_{n3}) \leq x(b) \leq \max(x'_{n1}, x'_{n3})
	\label{eq:check_x}
\end{equation}
\begin{equation}
	0 \leq z(b) \leq z'_{n}
	\label{eq:check_z}
\end{equation}
where $z'_n=z'_{n1}=z'_{n2}=z'_{n3}=z'_{n4}$.

Similarly, for a face with constant \( x = x'_{nj'} \), where $j'\in\{2,4\}$, we obtain:
\begin{equation}
	b = \frac{x'_{nj'} - x_{\text{u}}(t_i)}{x_i - x_{\text{u}}(t_i)}
	\label{eq:b_x}
\end{equation}
If \( b \) is within the interval \([0, 1]\), there is a possibility that the line segment intersects the face. Then, we compute:
\begin{equation}
	y(b) = (1 - b) y_{\text{u}}(t_i) + b y_i
	\label{eq:y_b}
\end{equation}
\begin{equation}
	z(b) = (1 - b) z_{\text{u}}(t_i) + b z_i
	\label{eq:z_b_2}
\end{equation}
and check if obtained \( y \) and \( z \) values are within the face boundaries:
\begin{equation}
	\min(y'_{n2}, y'_{n4}) \leq y(b) \leq \max(y'_{n2}, y'_{n4})
	\label{eq:check_y}
\end{equation}
\begin{equation}
	0 \leq z(b) \leq z'_{n}
	\label{eq:check_z_2}
\end{equation}
The above conditions were checked for a single cube. To examine the LoS between the UAV at \(\mathbf{r}_{u,t_i}\) and the user at \(\mathbf{r}_{i}\), these conditions must be verified for all $N_b$ cubes. If no intersections are found, the line segment is in LoS.

\begin{algorithm}
	\caption{Checking Line of Sight (LoS)}
	\label{alg:los_check}
	\begin{algorithmic}[1]
		\REQUIRE Coordinates of the UAV service point $(x_{\text{u}}(t_i), y_{\text{u}}(t_i), z_{\text{u}}(t_i))$ and the user $(x_i, y_i, z_i)$, coordinates of the four upper vertices of $N_b$ cuboids.
		\ENSURE LoS condition between the UAV service point and the user.
		
		\FOR {each cuboid $n = 1$ to $N_b$}
		\FOR {each face of the cuboid}
		\IF {face is in the $xz$-plane with constant $y = y'_{n1}$ or $y = y'_{n2}$}
		\STATE Compute $b$ using equation (\ref{eq:b_y})
		\IF {$0 \leq b \leq 1$}
		\STATE Compute $x(b)$ and $z(b)$ using equations (\ref{eq:x_b}) and (\ref{eq:z_b})
		\IF {$\min(x'_{n1}, x'_{n3}) \leq x(b) \leq \max(x'_{n1}, x'_{n3})$ and 
			$0 \leq z(b) \leq z'_{n}$}
		\RETURN Not in LoS
		\ENDIF
		\ENDIF
		\ELSIF {face is in the $yz$-plane with constant $x = x'_{n1}$ or $x = x'_{n2}$}
		\STATE Compute $b$ using equation (\ref{eq:b_x})
		\IF {$0 \leq b \leq 1$}
		\STATE Compute $y(b)$ and $z(b)$ using equations (\ref{eq:y_b}) and (\ref{eq:z_b_2})
		\IF {$\min(y'_{n1}, y'_{n2}) \leq y(b) \leq \max(y'_{n1}, y'_{n2})$ and 
			$0 \leq z(b) \leq z'_{n}$}
		\RETURN Not in LoS
		\ENDIF
		\ENDIF
		\ENDIF
		\ENDFOR
		\ENDFOR
		
		\RETURN In LoS
	\end{algorithmic}
\end{algorithm}

Finally, using these points, Algorithm \ref{alg:los_check} is presented. With this algorithm, for any UAV position in 3D space, we can quickly determine the LoS status for each user. Algorithm \ref{alg:los_check} first takes the coordinates of the UAV service point, the user, and the coordinates of the four upper vertices of all cuboids in the environment. For each cuboid, it checks each of its faces. If a face is in the \( xz \)-plane (constant \( y \)), it computes the parameter \( b \) using equation (\ref{eq:b_y}). If \( b \) is within the interval \([0, 1]\), it computes the corresponding \( x(b) \) and \( z(b) \) values using equations (\ref{eq:x_b}) and (\ref{eq:z_b}). It then checks if these values are within the boundaries of the face using equations (\ref{eq:check_x}) and (\ref{eq:check_z}). If they are, the line segment intersects the face, and the function returns ``Not in LoS". For faces in the \( yz \)-plane (constant \( x \)), it computes the parameter \( b \) using equation (\ref{eq:b_x}). If \( b \) is within the interval \([0, 1]\), it computes the corresponding \( y(b) \) and \( z(b) \) values using equations (\ref{eq:y_b}) and (\ref{eq:z_b_2}). It then checks if these values are within the boundaries of the face using equations (\ref{eq:check_y}) and (\ref{eq:check_z_2}). If they are, the line segment intersects the face, and the function returns ``Not in LoS". If no intersections are found after checking all faces of all cuboids, the function returns ``In LoS".

	\subsection{Quality of Service (QoS) Condition}
	In addition to LoS state, which is a necessary condition, the QoS must also be ensured during the trajectory.
	Based on \eqref{eq:ber_wi}, the QoS is quantified using BER, which depends on the user's risk level \( w_i \). Eq. \eqref{eq:ber_wi} ensures that users with higher risk levels \( w_i \) are provided with a better QoS, characterized by a lower BER.

	The BER of a UAV-based THz link is a function of various channel parameters. However, most of these parameters, such as carrier frequency, antenna pattern, and UAV's vibrations model and angular instability, remain constant throughout the trajectory \cite{dabiri2023enabling,dabiri2022pointing}. Besides LoS, which changes its status along the trajectory, the link length is the most important variable parameter affecting the QoS. Without losing generality and for simplicity, here, the BER is modeled using the \( Q \)-function, which is valid for most conditions.
	The BER can be approximated as:
	\begin{equation}
		\text{BER} \approx Q\left(\sqrt{\frac{2 \cdot P_r}{N_0}}\right)
		\label{eq:ber}
	\end{equation}
	where \( P_r \) is the received power, and \( N_0 \) is the noise power density. The received power \( P_r \) can be modeled as:
	\begin{equation}
		P_r = \frac{c_1 P_t}{d_{t_i}^2} \propto \frac{P_t}{d_{t_i}^2}
		\label{eq:pr}
	\end{equation}
	where \( P_t \) is the transmitted power and \( d_{t_i} \) is the distance between the UAV and the user:
	\begin{equation}
		d_{t_i} = \sqrt{(x_{\text{u}}(t_i) - x_i)^2 + (y_{\text{u}}(t_i) - y_i)^2 + (z_{\text{u}}(t_i) - z_i)^2}
		\label{eq:dti}
	\end{equation}
	Thus, the BER can be expressed as a function of distance:
	\begin{equation}
		\text{BER} \approx Q\left(\sqrt{\frac{2 \cdot c_1 \cdot P_t}{N_0 \cdot d_{t_i}^2}}\right).
		\label{eq:ber_distance}
	\end{equation}
	Based on \eqref{eq:ber_wi} and \eqref{eq:ber_distance}, to guarantee a BER below a certain threshold for a given \( w_i \), the distance \( d_{t_i} \) must satisfy:
	\begin{equation}
		d_{t_i} \leq \sqrt{\frac{2 \cdot c_1 \cdot P_t}{N_0 \cdot \left(Q^{-1}\left(\text{BER}_\text{th,1}^{(1 - w_i)} \cdot \text{BER}_\text{th,0}^{w_i}\right)\right)^2}}.
		\label{eq:dti_wi}
	\end{equation}

	%For a service point \( \mathbf{r}_{u,t_i} = (x_{\text{u}}(t_i), y_{\text{u}}(t_i), z_{\text{u}}(t_i)) \) to be valid for servicing the \(i\)-th user, the conditions provided in Algorithm 1 ($\text{LoS}_{i,t_i} = 1$) along with \eqref{eq:dti_wi} must be satisfied. 
	Using these results, Algorithm 2 is presented which determines a service point \( \mathbf{r}_{u,t_i} = (x_{\text{u}}(t_i), y_{\text{u}}(t_i), z_{\text{u}}(t_i)) \) to be valid based on the required QoS according to the user's risk level \( w_i \).

	\begin{algorithm}
		\caption{Validating Service Points for QoS}
		\label{alg:qos_validation}
		\begin{algorithmic}[1]
			\REQUIRE Coordinates of the UAV service point $(\mathbf{r}_{u,t_i})$ and the user $(\mathbf{r}_{i})$, user's risk level $w_i$, maximum and minimum acceptable BER $(\text{BER}_\text{th,1}, \text{BER}_\text{th,0})$, maximum distance $d_{\max}$, coordinates of the four upper vertices of $N_b$ cuboids.
			\ENSURE Validity of the service point for servicing the user.
			
			\STATE Compute the distance $d_{t_i}$ using equation (\ref{eq:dti})
			\STATE Calculate the required BER using equation (\ref{eq:ber_wi})
			\STATE Compute the maximum allowable distance $d_{t_i}$ for the given $w_i$ using equation (\ref{eq:dti_wi})
			
			\STATE Check the LoS condition using Algorithm \ref{alg:los_check}
			
			\IF {$d_{t_i}$ satisfies the distance condition in equation (\ref{eq:dti_wi}) and LoS is true}
			\RETURN Valid Service Point
			\ELSE
			\RETURN Invalid Service Point
			\ENDIF
			
		\end{algorithmic}
	\end{algorithm}

\subsection{Grid-based Evaluation of Service Area for User \( i \)}
Note that Algorithms 1 and 2 determine for a given point $(\mathbf{r}_{u,t_i})$ whether it guarantees the desired QoS for servicing $i$th user or not.
In this section, we extend the validation approach from a single service point \(\mathbf{r}_{u,t_i}\) to a service area \(\mathbf{S}_{u,i}\), which is a region parallel to the ground at a fixed altitude \( z_u(t) \). To this end, we discretize this area into a grid along the $x$ and $y$ directions and evaluate each grid point to determine if it can provide the required QoS for user \( i \) based on the conditions established in Algorithms 1 and 2.

To systematically evaluate the service area \(\mathbf{S}_{u,i}\), we perform the following steps. The service area \(\mathbf{S}_{u,i}\) is defined by its bounds \([x_{\text{min}}, x_{\text{max}}]\) and \([y_{\text{min}}, y_{\text{max}}]\) for any given altitude \( z_u(t) \). The area is discretized into a grid with a resolution of \(\Delta x\) and \(\Delta y\). The resulting grid is represented as a two-dimensional matrix \(\mathbf{S}_{u,i}\) with dimensions \( M_x \times M_y \), where:
\begin{align}
	\begin{cases}
		M_x = \left\lfloor \frac{x_{\text{max}} - x_{\text{min}}}{\Delta x} \right\rfloor + 1, \\
		%--------------
		M_y = \left\lfloor \frac{y_{\text{max}} - y_{\text{min}}}{\Delta y} \right\rfloor + 1. 
	\end{cases}
\end{align}
Each element of the matrix \(s_{u,i}(m_x, m_y)\) represents a grid point in the service area, where \( m_x \) and \( m_y \) are the indices of the matrix. The value of \(s_{u,i}(m_x, m_y)\) is defined as follows:
\begin{align}
	\begin{cases}
		s_{u,i}(m_x, m_y) = 1, ~\text{if the grid point satisfies Algorithm 2}\\
		%--------------
		s_{u,i}(m_x, m_y) = 0, ~ \text{otherwise}.
	\end{cases}
\end{align}

\begin{algorithm}
	\caption{Evaluating Service Area for User \( i \)} 
	\label{alg:service_area_evaluation}
	\begin{algorithmic}[1]
		\REQUIRE Bounds of the service area \([x_{\text{min}}, x_{\text{max}}]\) and \([y_{\text{min}}, y_{\text{max}}]\), grid resolution \(\Delta x, \Delta y\), fixed altitude \( z_u(t) \), user's coordinates \( \mathbf{r}_{i} = (x_i, y_i, z_i) \), user's risk level \( w_i \), maximum and minimum acceptable BER \( (\text{BER}_\text{th,1}, \text{BER}_\text{th,0}) \), maximum distance \( d_{\max} \), coordinates of the four upper vertices of \( N_b \) cuboids.
		\ENSURE Valid grid points for servicing the user.
		
		\STATE Initialize matrix \(\mathbf{S}_{u,i}\) with dimensions \( M_x \times M_y \)
		
		\FOR {each \( m_x \) from \( 0 \) to \( M_x - 1 \)}
		\FOR {each \( m_y \) from \( 0 \) to \( M_y - 1 \)}
		\STATE Set \( x = x_{\text{min}} + m_x \cdot \Delta x \)
		and \( y = y_{\text{min}} + m_y \cdot \Delta y \)
		\STATE Set \( \mathbf{r}_{u,t_i} \) by using $x$ and $y$
		
		\STATE Evaluate \( s_{u,i}(m_x, m_y) \) using Algorithm \ref{alg:qos_validation} 
		
		\IF {conditions in Algorithm \ref{alg:qos_validation} are satisfied}
		\STATE Set \( s_{u,i}(m_x, m_y) = 1 \)
		\ELSE
		\STATE Set \( s_{u,i}(m_x, m_y) = 0 \)
		\ENDIF
		\ENDFOR
		\ENDFOR
		
		\RETURN Matrix \(\mathbf{S}_{u,i}\) with valid grid points
	\end{algorithmic}
\end{algorithm}

%%%%%%%%%%%%%%%%%%%%%%%%%%%%%%%%%%%%%%%%%%%%%%%%%%%%%%%%%%%%%%%%
%%%%%%%%%%%%%%%%%%%%%%%%%%%%%%%%%%%%%%%%%%%%%%%%%%%%%%%%%%%%%%%% VERSUS W_Z
\begin{figure}
	\centering
	\subfloat[] {\includegraphics[width=3.4 in, trim=0cm 3cm 0cm 0cm, clip]{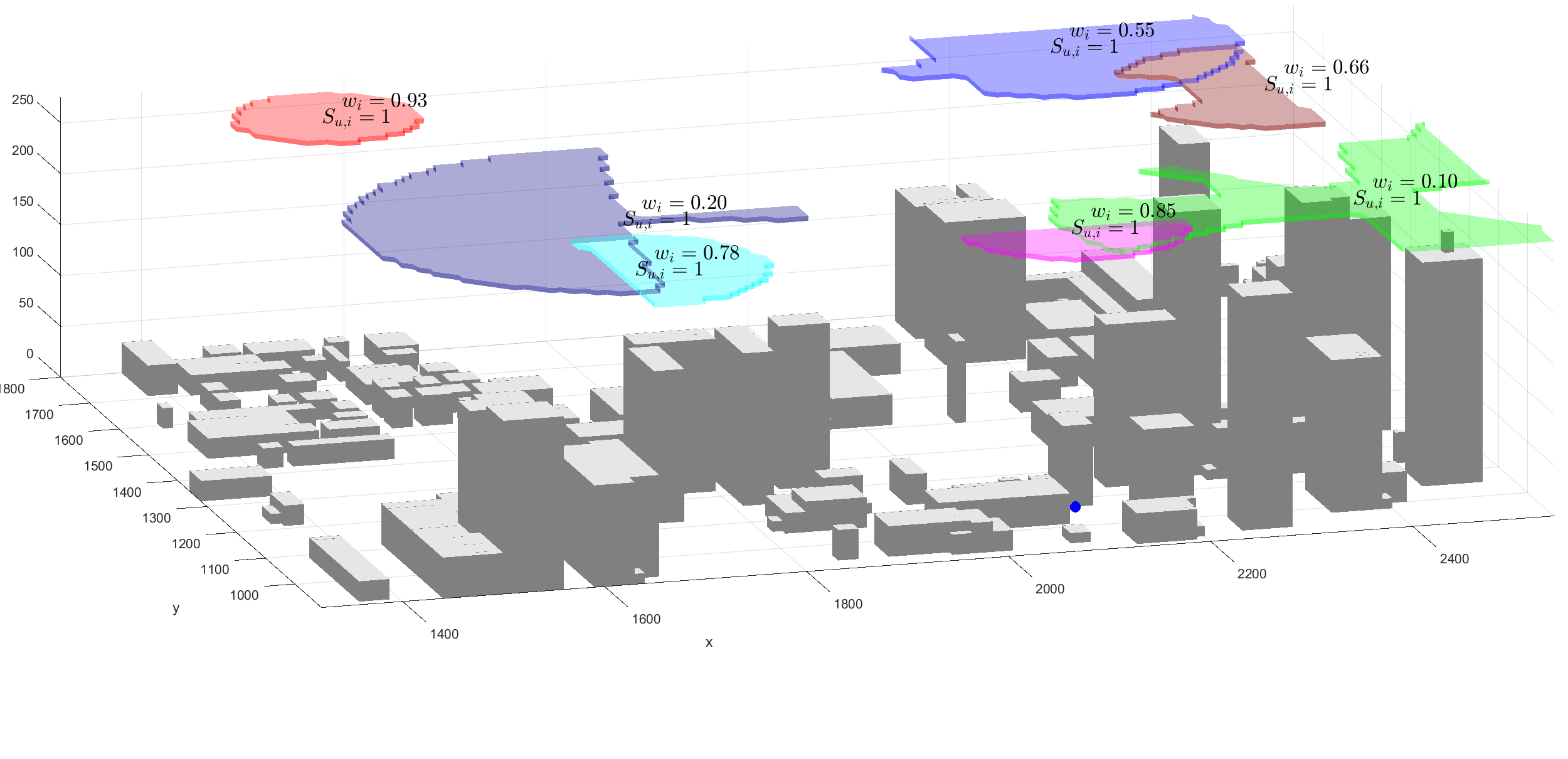}
		\label{ch1}
	}
	\hfill
	\subfloat[] {\includegraphics[width=3.2 in, height=2.1 in]{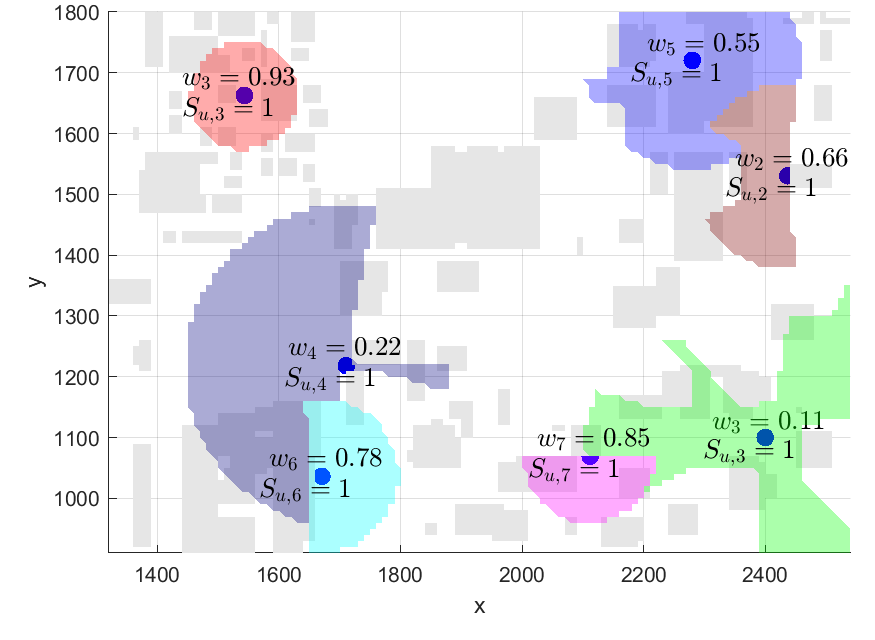}
		\label{ch2}
	}
	\caption{Illustration of the service area matrices \(\mathbf{S}_{u,i}\) for 7 users distributed in the West Bay area of Doha, Qatar, using Algorithm 3. The 3D view (a) and the 2D view (b) show the users' locations and their corresponding service areas. Each user is assigned a priority weight \(w_i\) and the service area \(\mathbf{S}_{u,i}\) is highlighted by the UAV at a given altitude of \(z_u(t) = 260\) m.}
	\label{ch}
\end{figure}
%%%%%%%%%%%%%%%%%%%%%%%%%%%%%%%%%%%%%%%%%%%%%%%%%%%%%%%%%%%%%%%%
%%%%%%%%%%%%%%%%%%%%%%%%%%%%%%%%%%%%%%%%%%%%%%%%%%%%%%%%%%%%%%%%  
%

Finally, utilizing the results from Algorithms 1 and 2, and after gridding the service area, Algorithm 3 is introduced. It rapidly calculates the matrix \( \mathbf{S}_{u,i} \) for any given altitude \( z_u(t) \) by loading data regarding the characteristics of 3D obstacles/buildings and also by determining the location of user \( i \) among the 3D obstacles.
To better understand, Fig. \ref{ch} illustrates the output of Algorithm 3 for a region in Doha, Qatar, specifically the West Bay area. For this purpose, data was downloaded from \cite{blender-osm} and, after applying a filter similar to that shown in Fig. \ref{cm2}, it was used as input for Algorithms 1, 2, and 3. Additionally, 7 users with randomly assigned locations and priority weights were generated as inputs for Algorithm 3. For a given altitude 
\( z_u(t)=260 \) m, the matrices \(\mathbf{S}_{u,i}\) for $i\in\{1,...,7\}$  have been produced, and the area corresponding to each user has been depicted in Fig. \ref{ch}.

Before starting the optimization process, we construct the matrix \(\mathbf{S}_{u,i}\) for all users, considering the physical characteristics of the environment and the users' locations. This matrix identifies the permissible locations for the UAV to provide service to user \( i \). Note that the matrix \(\mathbf{S}_{u,i}\) is obtained for a specific UAV altitude. If we aim to perform trajectory optimization in 3D, an additional dimension for altitude must be added to the matrix, and the same steps should be repeated for different altitudes.

%% -----------------------------
%% -----------------------------
\section{Optimization Problem Solution}
%% -----------------------------
%% -----------------------------

\subsection{Reformulation of the Optimization Problem}
Here, we reformulate the optimization problem defined earlier to address the inherent trade-off between minimizing the total mission time in \eqref{eq:objective_min_time} and prioritizing users based on their risk levels in \eqref{eq:priority_constraint}. The initial objective and this constraint often conflict with each other; prioritizing higher-risk users can lead to a longer overall service time. To manage this trade-off, we introduce a new objective function to balance the total service time and the priority of users as
\begin{align}
	\min \quad & \sum_{i=1}^{N_p} w_i^{I_w} \cdot t_i \label{eq:new_objective} \\
	%----
	\text{subject to} \quad & t_\text{end}= t_{N_p}+\frac{\|\mathbf{r}_{u,t_{N_p}} - \mathbf{r}_{\text{station}}\|}{v_{\text{UAV}}} \leq T_{\text{max}} \label{eq:simplified_energy_constraint} \\
	& \mathbf{r}_{u,t_i} \in \{\mathbf{S}_{u,i} \mid s_{u,i}(m_x, m_y) = 1 \} \label{eq:service_area_constraint}\\
	%-------------------------------------
	& \mathbf{r}_{u,0} = \mathbf{r}_{u,t_\text{end}} = \mathbf{r}_{\text{station}} \label{t_0} ,
\end{align}
where $v_{\text{UAV}}$ is the speed of UAV.
In the objective function \eqref{eq:new_objective}, the parameter \(I_w \geq 0\) allows us to control the trade-off between minimizing the total service time and giving priority to users with higher risk levels. When \(I_w = 0\), the objective function only minimizes the total service time without considering the priorities. As \(I_w\) increases, the importance of servicing higher priority users first becomes more significant.

The constraint \eqref{eq:simplified_energy_constraint} ensures that the total flight time $t_\text{end}$, including the time required to return to the charging station, does not exceed the maximum allowable flight time, \(T_{\text{max}}\), which is obtained as
\begin{align}
	\label{eq:t_max_relation}
	T_{\text{max}} = \frac{E_{\text{max}}}{P_{\text{UAV}}}
\end{align}
where \( P_{\text{UAV}} \) and \(E_{\text{max}}\) are the power consumption and the maximum energy capacity of the UAV, respectively. The term $\|\mathbf{r}_{u,t_{N_p}} - \mathbf{r}_{\text{station}}\|\big/v_{\text{UAV}}$ in \eqref{eq:simplified_energy_constraint} is the return time from the last service point \(\mathbf{r}_{u,t_{N_p}}\) to the charging station \(\mathbf{r}_{\text{station}}\).

Moreover, the constraint \eqref{eq:service_area_constraint} ensures the UAV's position \(\mathbf{r}_{u,t_i}\) at time \(t_i\) must be within the service area \(\mathbf{S}_{u,i}\) where the grid point is valid (i.e., \(s_{u,i}(m_x, m_y) = 1\)). 
%------------------------------
\subsection{Optimization Solution using Genetic Algorithm}
%------------------------------
Although the optimization problem \eqref{eq:new_objective} has been simplified as much as possible, it still has a large scale and a high computational complexity.
Given the complexity of the optimization problem, we first employ a Genetic Algorithm (GA) to find a near-optimal solution. GA is a metaheuristic inspired by the process of natural selection and is particularly suitable for solving complex optimization problems with large search spaces.

GA is chosen for this problem due to several reasons. Firstly, GAs are relatively straightforward to implement and do not require extensive knowledge about the problem domain, making them simple to use. Secondly, GAs perform well in large, complex search spaces where traditional optimization methods may struggle, showcasing their effectiveness in such scenarios. Additionally, GAs are highly flexible and can handle a variety of objective functions and constraints, making them suitable for the multi-objective nature of this problem. Lastly, the stochastic elements of GAs, such as selection, crossover, and mutation, help in exploring a wide range of solutions, thereby reducing the risk of getting trapped in local optima.

The GA-based method is summarized in Algorithm 4 to solve our optimization problem.
The fitness function evaluates how well each solution satisfies the objective function and constraints. For our problem, the fitness function \( F \) can be defined as follows:
\begin{align}
	F &= -\left( \sum_{i=1}^{N_p} w_i^{I_w} \cdot t_i \right) + \lambda \left( T_{\text{max}} - t_\text{end} \right) \nonumber \\
	& + \mu \sum_{i=1}^{N_p} \mathbf{1}\left( \mathbf{r}_{u,t_i} \in \mathbf{S}_{u,i} \right) - \gamma \mathbf{1}\left( \mathbf{r}_{u,t_\text{end}} \neq \mathbf{r}_{\text{station}} \right)
\end{align}
where \( \lambda \) and \( \mu \) are penalty factors, and \( \mathbf{1} \) is an indicator function that ensures the service area constraint is satisfied for all users.

\begin{algorithm}
	\caption{Genetic Algorithm for UAV Trajectory Optimization}
	\label{alg:ga_uav}
	\begin{algorithmic}[1]
		\REQUIRE Population size \( P \), maximum generations \( G \), crossover rate \( p_c \), mutation rate \( p_m \)
		\STATE Initialize a population of size \( P \) with random solutions
		\FOR{generation = 1 to \( G \)}
		\STATE Evaluate the fitness \( F \) of each solution in the population
		\STATE Select parents based on fitness values
		\STATE Apply crossover with probability \( p_c \) to produce offspring
		\STATE Apply mutation with probability \( p_m \) to offspring
		\STATE Evaluate the fitness \( F \) of the new offspring
		\STATE Select the next generation from the current population and offspring
		\ENDFOR
		\STATE \textbf{return} The best solution found
	\end{algorithmic}
\end{algorithm}

The Algorithm 4 proceeds through several stages. Firstly, the initialization stage involves generating an initial population of random solutions, each representing a possible UAV trajectory. Next, in the evaluation stage, the fitness of each solution is assessed using the fitness function, which incorporates the objective function and constraints. During the selection stage, solutions are chosen based on their fitness values, with higher fitness solutions having a greater chance of being selected. The crossover stage involves combining pairs of solutions to produce new offspring, introducing new solutions into the population. In the mutation stage, random changes are applied to some offspring to maintain diversity and explore new areas of the solution space. The replacement stage forms the new generation by selecting the best solutions from the current population and the new offspring. Finally, the termination stage repeats this process until the maximum number of generations is reached or a satisfactory solution is found. The service area matrix \( \mathbf{S}_{u,i} \) for each user is obtained based on Algorithm 3, ensuring that both QoS and LoS constraints are satisfied for each user's location.

%-----------------------------------
\subsection{Enhanced Heuristic Algorithm for Faster Convergence}
%------------------------------------
Although Algorithm 4 effectively solves the problem, the large scale of the problem results in a prolonged convergence time to the optimal solution. Therefore, in this section, we propose a hybrid heuristic algorithm based on GA and environmental physical information, which converges to the optimal solution much faster.

To address the problem, we present a method where initially, for each user \(i\), a point is selected randomly from the valid service area matrix \(\mathbf{S}_{u,i}\) as
\begin{equation}\label{p1}
	\mathbf{r}_{u,t_i} \in \{\mathbf{r}_{u,t} \mid s_{u,i}(m_x, m_y) = 1\},~~i\in\{1,...,N_p\}
\end{equation}
Then, the trajectory optimization problem is formulated to find the optimal path for the \(N_p\) selected points to minimize the objective function. After solving this problem, the method is repeated multiple times, where each time a random point is selected for each user, and the optimization problem is solved to minimize the objective function.

After selecting each point $\mathbf{r}_{u,t_i}$ according to \eqref{p1}, the optimization problem is reduced to selecting the minimum time.
Let $i$ denote the next user to receive service after the current user $j$. The travel time between points $\mathbf{r}_{u,t_i}$ and $\mathbf{r}_{u,t_j}$ is defined as:
\begin{equation}
	t_i = t_j + \sum_{j=0}^{N_p} t_{ji} \cdot a_{ji}, \quad \forall i
	\label{eq:travel_time_consistency}
\end{equation}
where
\begin{align}
	t_{ji} = \frac{\|\mathbf{r}_{u,t_i} - \mathbf{r}_{u,t_j} \|}{v_{\text{UAV}}}, \quad \forall j, i
	\label{eq:travel_time}
\end{align}
and
\begin{equation}
	a_{ji} = 
	\begin{cases} 
		1, & \text{if the UAV travels from}~\mathbf{r}_{u,t_j} \text{ to}~ \mathbf{r}_{u,t_i} \\
		&\text{ immediately after } ~\mathbf{r}_{u,t_j}; \\
		0, & \text{otherwise}.
	\end{cases}
	\label{eq:a_ji_definition}
\end{equation}
The constraint \eqref{eq:travel_time_consistency} ensures that the service time for user \( i \) is equal to the service time for user \( j \) plus the travel time between points \( j \) and \( i \). This is necessary to ensure a logical and orderly sequence of visits and accurate computation of service times.

To visit each randomly selected position $\mathbf{r}_{u,t_i}$ only once, we have to:
\begin{align}
	& \sum_{j=0}^{N_p} a_{ji} = 1, \quad \forall i
	\label{eq:visit_each_position_once_1} \\
	%-------
	& \sum_{i=0}^{N_p} a_{ij} = 1, \quad \forall j
	\label{eq:visit_each_position_once_2}.
\end{align}
These constraints \eqref{eq:visit_each_position_once_1} and \eqref{eq:visit_each_position_once_2} ensure that each selected point is visited exactly once, and the UAV returns to the starting position. This is essential to ensure that all users are served without repetition and that the UAV completes its mission efficiently.

In order for the trajectory path to start from the charging station and return to the charging station at the end of the path, the following conditions must also be met:
\begin{align}
	& \sum_{i=1}^{N_p} a_{0i} = 1
	\label{eq:start_at_starting_position} \\
	%-------------
	& \sum_{i=1}^{N_p} a_{i0} = 1
	\label{eq:end_at_starting_position}
\end{align}
These constraints \eqref{eq:start_at_starting_position} and \eqref{eq:end_at_starting_position} ensure that the UAV starts from the initial position and returns to it after serving all users. This is important for practical implementation, ensuring that the UAV can recharge and prepare for subsequent missions.

To prevent sub-tours, the following constraints are included:
\begin{align}
	& u_i - u_j + N_p \cdot a_{ji} \leq N_p - 1, \quad \forall 2 \leq i, j \leq N_p, \, i \neq j
	\label{eq:subtour_elimination_1} \\
	%--------------------------------
	& u_i \geq 2, \, u_i \leq N_p, \quad \forall 2 \leq i \leq N_p
	\label{eq:subtour_elimination_2},
\end{align}
where $u_1 = 1$. These constraints \eqref{eq:subtour_elimination_1} and \eqref{eq:subtour_elimination_2} prevent the formation of sub-tours, which are smaller loops that do not include all users. Sub-tours can lead to inefficient paths and increased mission time. Eliminating sub-tours ensures that the UAV follows a single tour that visits all users once.

Using the above points, we propose an enhanced heuristic algorithm that combines random point selection and an optimization framework for faster convergence. This algorithm, detailed in Algorithm \ref{alg:heuristic_optimization_detailed}, operates as follows. First, for each user \(i\), a point \(\mathbf{r}_{u,t_i}\) is randomly selected from the valid service area matrix \(\mathbf{S}_{u,i}\), which ensures that the QoS constraints are satisfied. This selection is repeated for a predefined number of iterations \(I\).

During each iteration, the algorithm initializes a list of selected points and formulates an optimization problem to find the optimal path connecting these points while minimizing the total mission time. The optimization problem includes constraints to ensure that each user is visited exactly once, the UAV starts and ends at the charging station, and sub-tours are eliminated.

The objective function in this algorithm balances minimizing the total service time and prioritizing users based on their risk levels. The solution from each iteration is evaluated, and the best solution is stored. After all iterations, the algorithm returns the best trajectory found, which minimizes the overall mission time while adhering to all constraints.

\begin{algorithm}
	\caption{Enhanced Heuristic Algorithm for Faster Convergence}
	\label{alg:heuristic_optimization_detailed}
	\begin{algorithmic}[1]
		\REQUIRE Number of iterations \( I \), service area matrices \(\mathbf{S}_{u,i}\) for all users, UAV speed \( v_{\text{UAV}} \), maximum allowable flight time \( T_{\text{max}} \), number of users \( N_p \), coordinates of charging station \( \mathbf{r}_{\text{station}} \)
		\STATE Initialize the best objective value \( F_{\text{best}} \) to a very large number
		\STATE Initialize the best solution \( \text{Sol}_{\text{best}} \) to an empty list
		\FOR{iteration = 1 to \( I \)}
		\STATE Initialize the list of selected points \( \mathbf{R}_\text{selected} \) to an empty list
		\FOR{each user \( i \) from \( 1 \) to \( N_p \)}
		\STATE Randomly select a valid point \( \mathbf{r}_{u,t_i} \) from \( \mathbf{S}_{u,i} \)
		\STATE Add \( \mathbf{r}_{u,t_i} \) to \( \mathbf{R}_\text{selected} \)
		\ENDFOR
		\STATE Initialize the optimization model
		\STATE Define the decision variables \( a_{ji} \) for all \( j, i \)
		\STATE Define the auxiliary variables \( u_i \) for all \( i \)
		\STATE Set the objective function as \( \min \sum_{i=1}^{N_p} w_i^{I_w} \cdot t_i \)
		\STATE Add the constraint \( t_i = t_j + \sum_{j=0}^{N_p} t_{ji} \cdot a_{ji} \) for all \( i \)
		\STATE Add the constraints \( \sum_{j=0}^{N_p} a_{ji} = 1 \) and \( \sum_{i=0}^{N_p} a_{ij} = 1 \) for all \( i, j \)
		\STATE Add the constraints \( \sum_{i=1}^{N_p} a_{0i} = 1 \) and \( \sum_{i=1}^{N_p} a_{i0} = 1 \)
		\STATE Add the subtour elimination constraints \( u_i - u_j + N_p \cdot a_{ji} \leq N_p - 1 \) for all \( 2 \leq i, j \leq N_p \), \( i \neq j \)
		\STATE Solve the optimization problem to find the optimal path \( \mathbf{R}_\text{optimal} \)
		\STATE Calculate the objective value \( F_\text{current} \) for the current solution
		\IF{\( F_\text{current} < F_\text{best} \)}
		\STATE Update \( F_\text{best} = F_\text{current} \)
		\STATE Update \( \text{Sol}_{\text{best}} = \mathbf{R}_\text{optimal} \)
		\ENDIF
		\ENDFOR
		\STATE \textbf{return} The best solution found \( \text{Sol}_{\text{best}} \)
	\end{algorithmic}
\end{algorithm}

%-----------------------------------
%-----------------------------------
\subsection{Advanced Heuristic Algorithm for Faster Convergence}
%-----------------------------------
%-----------------------------------
To further enhance the convergence speed of the optimization problem, we propose an advanced heuristic algorithm that builds upon the previous heuristic method in Algorithm 5 by incorporating user priority weights and optimizing the trajectory for higher-priority nodes first. Initially, for each user \(i\), a point \(\mathbf{r}_{u,t_i}\) is selected from the valid service area matrix \(\mathbf{S}_{u,i}\). Based on the distance constraint in equation \eqref{eq:dti_wi}, where users with lower priority weights (\(w_i\)) have larger allowable distances (\(d_{t_i}\)), lower priority users can initially be excluded. A random subset of lower-priority nodes is selected for potential exclusion. The number of randomly select nodes from the set of lower priority nodes is denoted by \(N'_p\), where \(N'_p\) is a random number from the set \(\{0, 1, \ldots, \frac{N_p}{2}\}\). The randomness of \(N'_p\) ensures that all possible scenarios are examined over numerous iterations.

After excluding some lower-priority nodes, the heuristic algorithm in Algorithm 5 is applied to the remaining $N''_p=(N_p-N'_p)$ higher-priority nodes. Once an optimal trajectory for the higher-priority nodes is found, the service area matrices \(\mathbf{S}_{u,i}\) of the excluded lower-priority nodes are checked. If there is an intersection with the optimized trajectory due to larger \(d_{t_i}\), these lower-priority nodes can be reintegrated into the trajectory. This process ensures that lower-priority nodes receive service without significantly increasing the total mission time. The algorithm repeats this process multiple times with different random selections of \(N'_p\) nodes to find the most optimal trajectory that includes as many nodes as possible. By strategically removing and potentially reintegrating lower-priority nodes, this advanced heuristic algorithm aims to find a more optimal trajectory faster while ensuring service to as many nodes as possible (Algorithm \ref{alg:advanced_heuristic_optimization}).

\begin{algorithm}
	\caption{Advanced Heuristic Algorithm for Faster Convergence}
	\label{alg:advanced_heuristic_optimization}
	\begin{algorithmic}[1]
		\REQUIRE Number of iterations \( I \)
		\FOR{iteration = 1 to \( I \)}
		\STATE Randomly select a point \(\mathbf{r}_{u,t_i}\) for each user \(i\) from \(\mathbf{S}_{u,i}\)
		\STATE Randomly select \(N'_p\) nodes from the lower priority nodes, where \(N'_p \in \{0, 1, \ldots, \frac{N_p}{2}\}\)
		\STATE Remove the selected \(N'_p\) nodes from the optimization problem
		\STATE Apply the heuristic algorithm \ref{alg:heuristic_optimization_detailed} to the remaining nodes
		\STATE Check if the excluded lower priority nodes can be reintegrated based on the optimized trajectory and distance constraints
		\IF{possible to reintegrate all excluded nodes}
		\STATE Store the solution and the corresponding objective value
		\ENDIF
		\ENDFOR
		\STATE \textbf{return} The best solution found across all iterations
	\end{algorithmic}
\end{algorithm}

%%%%%%%%%%%%%%%%%%%%%%%%%%%%%%%%%%%%%%%%%%%%%%%%%%%%%%%%%%%%%%%%
%%%%%%%%%%%%%%%%%%%%%%%%%%%%%%%%%%%%%%%%%%%%%%%%%%%%%%%%%%%%%%%% VERSUS P_T
\begin{figure}
	\begin{center}
		\includegraphics[width=3.3 in]{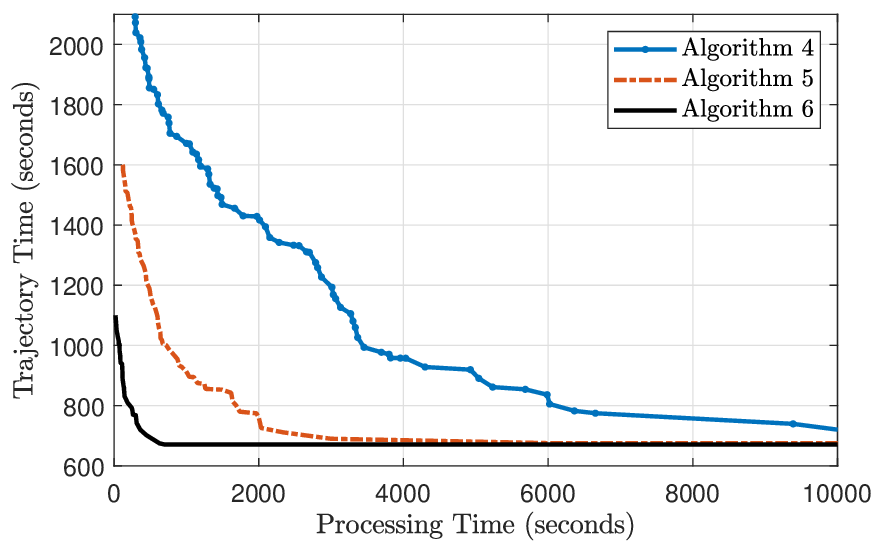}
		\caption{Comparison of the performance of Algorithm 4, Algorithm 5, and Algorithm 6 in terms of trajectory time versus processing time for \(N_p = 10\) and \(I_w = 2\).}
		\label{sg1}
	\end{center}
\end{figure}

\begin{figure*}
	\begin{center}
		\includegraphics[width=7.2 in, trim=0cm 5cm 0cm 0cm, clip]{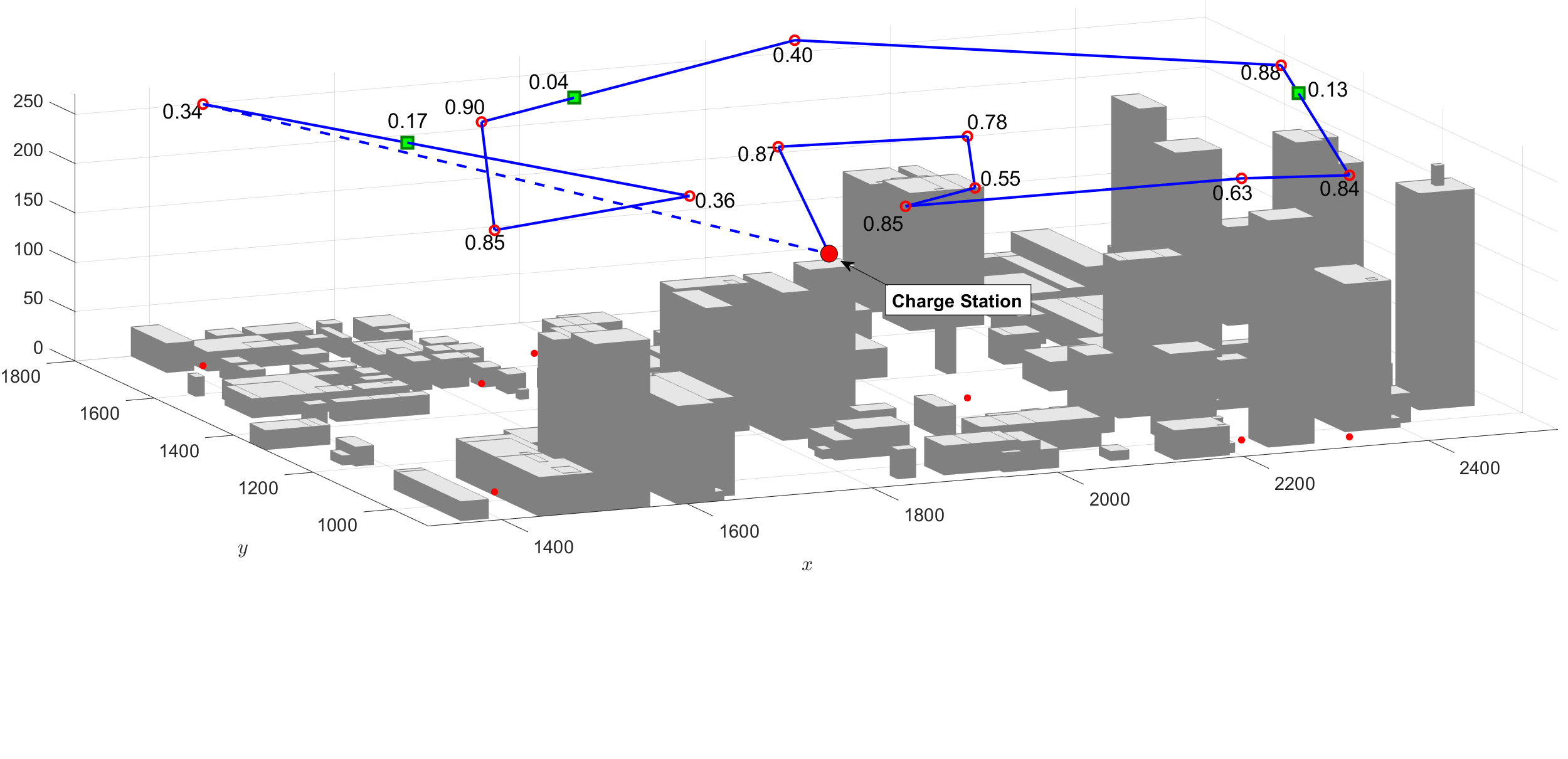}
		\caption{Trajectory output of Algorithm 6 for \(I_w = 2\) with 14 users randomly distributed among 3D obstacles. The UAV starts from the charging station, services the users, and returns to the charging station following the dashed path. Red circles indicate points where the UAV provides or ensures service quality to distributed nodes. Green squares represent locations where lower-weight nodes are serviced without affecting the overall trajectory time.}
		\label{sf1}
	\end{center}
\end{figure*}

%% --------------------------------------
%% --------------------------------------
\section{Simulation Results}
%% --------------------------------------
%% --------------------------------------
In this section, we aim to evaluate the performance of the trajectory algorithms on real-world environment data. For this purpose, we extracted 3D data related to the West Bay area of Doha, Qatar from the database \cite{blender-osm}. After applying a filtering step similar to Fig. \ref{cm2}, we randomly distribute \(N_b\) users among the 3D obstacles. Each user is assigned a random priority weight \(w_i\). Additionally, we consider the UAV charging station to be located on one of the buildings in the target area. The UAV flight altitude is set to 10 meters above the height of the tallest building in the target area, i.e., \(z_u(t) = 260\) m. The threshold BER for \(w_i = 0\) is set to \(\text{BER}_\text{th,0}=10^{-3}\), and the threshold BER for \(w_i = 1\) is set to \(\text{BER}_\text{th,1}=10^{-6}\).

We also consider the UAV is equipped with a $10\times10$ THz antenna array operating at a carrier frequency of 140 GHz. Theoretically, this $10\times10$ array is expected to have dimensions of approximately $9.63\times 9.63$ mm$^2$, with each element spaced around 1.07 mm apart. The theoretical gain of this antenna array is approximately 15.03 dB \cite{balanis2016antenna}.
Additionally, the ground user antenna is considered as a $5\times5$ array with dimensions of approximately $4.8\times 4.8$ mm$^2$ mm and a theoretical gain of approximately 7.03 dB. The transmit power is set to 10 mW.

For the UAV, we considered the DJI Matrice 600 Pro, which is capable of carrying higher payloads required for telecommunications equipment. The Matrice 600 Pro can carry a payload of up to 6 kg and has a maximum flight time of approximately $T_\text{max}=2100$ seconds (35 minutes) with a fully charged battery \cite{DJIM600Pro}. At a constant speed of \(v_\text{uav} = 5\) m/s, this flight duration allows for significant operational time within the constraints of the scenario.

In Fig. \ref{sg1}, we compare the performance of three different algorithms in optimizing the UAV trajectory time as a function of processing time. The results are obtained for \(N_p = 10\) and \(I_w = 2\). The plot illustrates that Algorithm 6 converges to the optimal solution the fastest, achieving the shortest trajectory time in the least processing time. Algorithm 5 also converges relatively quickly but not as efficiently as Algorithm 6. In contrast, Algorithm 4 requires significantly more processing time to reach a similar trajectory time. The significant reduction in optimization search space in Algorithms 5 and 6 is achieved through the introduction of heuristic methods. In Algorithm 6, the additional heuristic step of randomly excluding lower-weight nodes further reduces the search space. This approach enables Algorithm 6 to find the optimal trajectory in a shorter time, making it highly practical for dynamic environments. In dynamic scenarios, the positions of users, the number of users, or their priority weights may change frequently. Hence, the quick convergence of Algorithm 6 makes it well-suited for dynamic scenarios.

To better understand the output of Algorithm 6, Fig. \ref{sf1} shows a trajectory example for \(I_w = 2\) where 14 users are randomly distributed among 3D obstacles. The UAV begins its trajectory from a charging station and, after providing service, returns to the charging station via the dashed path. The red circular points indicate locations where the UAV meets or ensures the quality of service for the distributed nodes. The main idea behind Algorithm 6 compared to Algorithm 5 is the random exclusion of lower-weight nodes, followed by trajectory optimization with the remaining higher-weight nodes. If the conditions of Algorithm 2 are met for the excluded points, they are reinserted into the trajectory. The green square points in the trajectory represent locations where the UAV services these lower-weight nodes. As observed, the lower-weight nodes are serviced along the trajectory path of two higher-priority nodes, effectively not impacting the overall trajectory time.

\begin{figure}
	\centering
	\subfloat[] {\includegraphics[width=3.3 in, height=1.9 in]{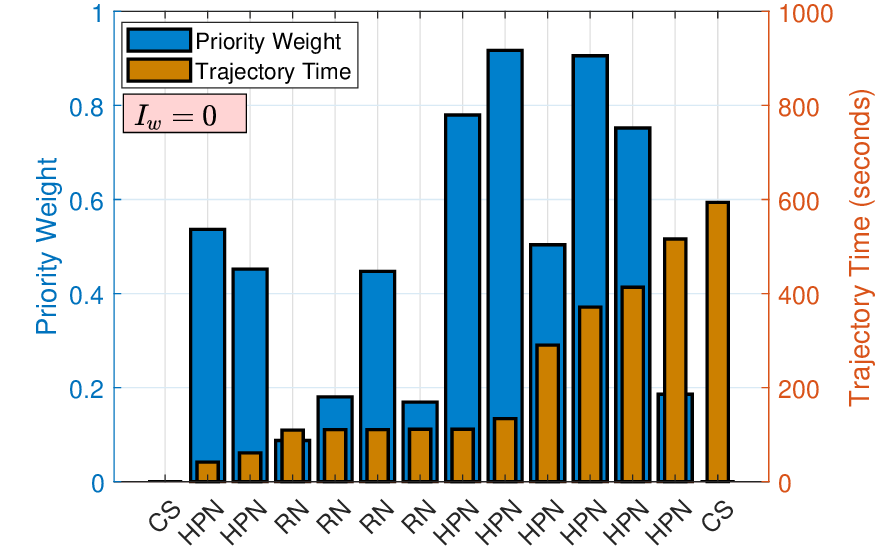}
		\label{cf1}
	}
	\hfill
	\subfloat[] {\includegraphics[width=3.3 in, height=1.9 in]{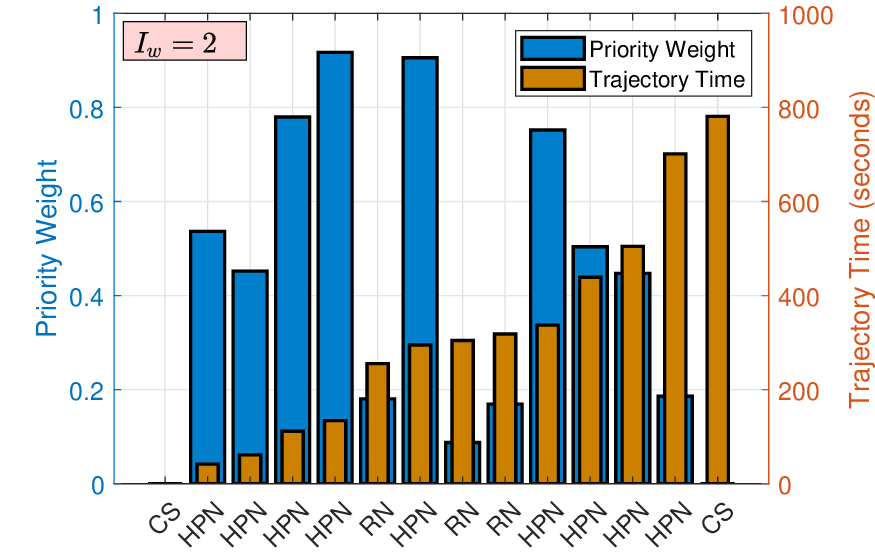}
		\label{cf2}
	}
	\hfill
	\subfloat[] {\includegraphics[width=3.3 in, height=1.9 in]{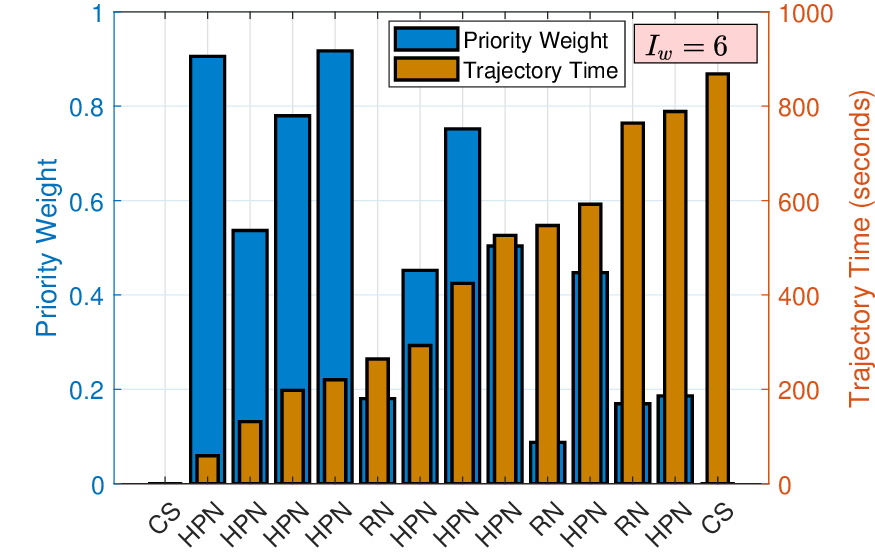}
		\label{cf3}
	}
	\caption{Comparison of trajectory times for different priority weight factors \(I_w\). (a) \(I_w = 0\), (b) \(I_w = 2\), and (c) \(I_w = 4\). Each subfigure illustrates the trajectory time for 12 users using Algorithm 6, starting from and returning to the charging station. The labels 'CS', 'RN', and 'HPN' indicate the charging station, removed nodes, and higher-priority nodes, respectively.}
	\label{cf}
\end{figure}

Figure \ref{cf} illustrates the impact of different priority weight factors \(I_w\) on the trajectory time for 12 users, using Algorithm 6. The subfigures show the trajectory starting from the charging station (CS), considering removed nodes (RN), and higher-priority nodes (HPN) for three different values of \(I_w\)= 0, 2, and 4. 
For \(I_w = 0\) (Fig. \ref{cf1}), the UAV prioritizes finding the shortest path to service all users, disregarding the priority weights of the nodes. This results in a trajectory time of approximately 600 seconds. The objective function \eqref{eq:new_objective} here emphasizes minimizing the total mission time without considering user priorities.

As \(I_w\) increases to 2 (Fig. \ref{cf2}), the UAV starts giving higher priority to nodes with higher weights, as dictated by the objective function \eqref{eq:new_objective}. This prioritization results in an increased trajectory time compared to \(I_w = 0\). The UAV adjusts its path to ensure that higher-priority nodes are serviced first, demonstrating the trade-off between mission time and priority weights.
With \(I_w = 4\) (Fig.e \ref{cf3}), the trend continues, and the UAV further prioritizes higher-weight nodes, leading to an even longer trajectory time. The increased \(I_w\) significantly influences the UAV's path, ensuring that the highest priority nodes are serviced first, even if it means a longer overall mission time. This demonstrates how the objective function \eqref{eq:new_objective} balances the trade-off between minimizing total service time and prioritizing higher-weight nodes.

%% ----------------------------------
%% ----------------------------------
\section{Conclusion}
%% ----------------------------------
%% ----------------------------------
This paper presented a comprehensive approach to optimizing UAV trajectories for emergency response operations in real 3D urban environments. We addressed the critical challenges of ensuring LoS connections, minimizing energy consumption, and prioritizing user service based on urgency during disaster recovery scenarios. We proposed a detailed modeling of 3D obstacles using online map data, formulating an optimal trajectory problem that integrates user priority levels and LoS constraints, and developing both a GA-based solution and an enhanced heuristic algorithm for faster convergence. The effectiveness of our proposed methods was demonstrated through simulations using real-world data from the West Bay area of Doha, Qatar. The results showed significant improvements in communication efficiency and reliability in disaster-stricken areas. By providing prioritized and reliable communication services, UAVs can play a pivotal role in modern wireless networks, particularly during critical emergency situations.


\end{document}